\documentclass[a4paper,11pt]{article}
\usepackage{jheppub} 
\usepackage[T1]{fontenc} 

\title{Deformation quantization in the teaching of Lie group representations}

\author[a]{Alexander J. Balsomo}
\author[b]{and Job A. Nable}

\affiliation[a]{West Visayas State University,\\Luna St. La Paz, Iloilo City, Philippines}
\affiliation[b]{Ateneo de Manila University,\\Katipunan Avenue, Loyola Heights, Quezon City, Philippines}

\emailAdd{abalsomo@wvsu.edu.ph}
\emailAdd{jnable@ateneo.edu}

\abstract{In this work, we present straightforward and concrete computations of the unitary irreducible representations of the Euclidean motion group $M(2)$ employing the methods of deformation quantization.  Deformation quantization is a quantization method of classical mechanics and is an autonomous approach to quantum mechanics, arising from the Wigner quasiprobability distributions and Weyl correspondence.  We advertise the utility and power of deformation theory in Lie group representations.  In implementing this idea, many aspects of the method of orbits is also learned, thus further adding to the mathematical toolkit of the beginning graduate student of physics.  Furthermore, the essential unity of many topics in mathematics and physics (such as Lie groups and Lie algebras, quantization, functional analysis and symplectic geometry) is witnessed, an aspect seldom encountered in textbooks, in an elementary way.}


\begin{document} 
\maketitle
\flushbottom

\section{Introduction}

In this paper, the deformation quantization approach to the representation theory of Lie groups is discussed via the example of the Euclidean motion group $M(2),$ which is the group of rigid motions of the plane. Representations of Lie groups are subsumed under the general theory of group representations. The development of quantum theory in the mid-1920s, with the appearance of von Neumann's book on the mathematical foundations of quantum mechanics, greatly influenced the theory of unitary representations of groups in infinite dimensional Hilbert space \cite{Neumann}. Succeeding early works in this field were due to Bargmann, Wigner, Gelfand-Naimark, and others \cite{Mackey}. The basic idea is as follows. Suppose a Lie group $G$ acts on a set $X$, denoted by $(g, x)\to g\cdot x.$ Let $V$ be a vector space of complex-valued functions on $X$ which is invariant under the action of $G$, that is, the function $x\to f(g\cdot x)$ is in $V$, whenever $f$ is in $V$. Thus, the mapping $T_g:f\to T_gf,$ where $(T_gf)(x)=f(g\cdot x)$ is a linear transformation on $V$ and is invertible. The mapping $g\to T_g$ from $G$ into the group $GL(V)$ of invertible linear transformations of $V$ is called a linear representation of $G$ in $V$.

Mechanics provides basic examples of group representations. In classical mechanics, an observable is a function on phase space $M$, which is a Poisson manifold, while in quantum mechanics, observables are self-adjoint operators on a Hilbert space. Quantization, as generally understood, is a mapping from classical observables to the space of quantum observables, where this mapping satisfies certain conditions first laid out formally by von Neumann. In the simplest case of the free particle, Dirac's canonical quantization of phase space variables turns out to be a representation of the Heisenberg Lie algebra, and the exponentiation of this representation gives the representation of the Heisenberg Lie group. This basic example already illustrates the deep and beautiful connections between quantization and representations of Lie groups. More generally, in the above definition of a linear representation, the classical observables are the functions $f$ on $M$ and $G$ acts on $M$. This induces an action of the Lie algebra of $G$ on the classical observables via vector fields. Modulo many technical difficulties, resolved in many general cases by the orbit method or geometric quantization of Kirillov \cite{Kirillov2}, Kostant \cite{Kostant} and Souriau \cite{Souriau}, the exponentiation of the Lie algebra representations give the quantum observables.

There are, currently, three accepted quantization procedures in quantum theory \cite{Zachos}. There is the canonical quantization developed earliest by Heisenberg, Schrodinger and others in the 1920s, the path integral method by Dirac and Feynman, and the phase space formulation of quantum mechanics or deformation quantization, which this work focuses on.

Phase space quantum mechanics is based on Wigner's quasiprobability distribution \cite{Wigner} and the Weyl correspondence \cite{Weyl} between self-adjoint operators in Hilbert space and ordinary functions, called the symbols of the operators. It turns out that the Weyl symbol of the projection onto a state is the Wigner function corresponding to the state. The Wigner function, which is a function on phase space, allows for the computation of quantum averages by classical like formulas. Moreover, its marginal distributions produce the correct probability distributions for the position and momentum of the system \cite[Appendix  A]{Hug}. Not least of its utility is that it is the approach that gives most insight into the connection between classical mechanics and quantum mechanics. It was Groenewold \cite{Groenewold} and Moyal \cite{Moyal} who first gave the formulas for the symbols of the composition and commutators of two quantum observables, now known as the Moyal star-product. In the early 1970s, Bayen et al. \cite{Bayen,Bayen1} elevated this formula as a definition of deformation of functions on Poisson manifolds and proposed deformation quantization as an autonomous quantum theory.

The central idea of deformation quantization is the deformation of the usual pointwise commutative product of functions on Poisson manifolds into a noncommutative and associative star-product or $\star-$product, and the deformation of the Poisson bracket arising from the associativity of the $\star-$product. In their seminal work, Bayen et al. suggested that quantization should be "a deformation of the structure of the algebra of classical observables and not as a radical change in the nature of the observables" \cite[p. 62]{Bayen}. Deformation quantization is a synthesis of works due to Weyl, Wigner, Moyal, Groenewold, Gerstenhaber, and others. In 1997, Kontsevich \cite{Kontsevich} proved the existence of deformation quantization of regular Poisson manifolds. Previous to this, Fedosov, in the early 1980s, gave a very nice geometric proof of the existence of deformation quantization of symplectic manifolds (originally found in \cite{Fedosov}, but later extended in \cite{Fedosov1}) and started the great interest on deformation quantization among mathematicians.

As a quantization theory, it is inevitable that deformation quantization found use into the representation theory of Lie groups. This has already been strongly hinted at in \cite{Bayen,Bayen1}. Subsequent developments in the works \cite{Arnal1,Arnal2,Arnal4,Arnal3,Arnal7,Arnal5,Fronsdal,Moreno} have shown that deformation theory, together with the orbit method, is very useful in representation theory. As the beautiful papers \cite{Hirshfeld}, from which we copied our title, and \cite{Case} have the aim of introducing deformation quantization and phase space methods in physics instruction, in particular in quantum mechanics, we also deemed it worthwhile to teach Lie group representations via the method of deformation quantization. In as much as \cite{Arnal1,Arnal2,Arnal4,Arnal3,Arnal7} have already attempted to use star-products in the representation theory of various classes of Lie groups, these papers assume many deep mathematical results and large gaps in the computations make them very difficult reading for beginning graduate students.

In this article, we present fairly complete and concrete computations in obtaining the irreducible unitary representations of a particular Lie group using deformation quantization. Works similar to our own are \cite{Diep1,Diep2,Nable,Nguyen}.  We suggest to readers Berndt's introductory text on symplectic manifold \cite{Berndt} or Abraham and Marsden's work \cite{Abraham} which is a more advanced approach.  Kirillov's orbit method \cite{Kirillov2} and introductory books in unitary representations by Sugiura \cite{Sugiura}, Berndt \cite{Berndt1} and Mackey \cite{Mackey1} are highly recommended.  In section \ref{UnitaryRepresentation} important concepts about unitary representations will be discussed, in particular, its construction by the method of induced representation and we also present the unitary representations of the Euclidean motion group $M(2)$. We will formally discuss quantization in section \ref{Quantization}.  The non-Hilbert space-based quantization, deformation quantization, the concept of $\star$-product and its connection to unitary representation theory will be discussed in section \ref{DeformationQuantization}.  In section \ref{UIRM(2)}, our main contribution is the concrete computation of the unitary representations of $M(2)$ via deformation quantization.  Finally, we summarize our results in section \ref{Conclusion}.

\section{Unitary Representations}\label{UnitaryRepresentation} 

A representation of a group $G$ on a vector space $V$ over a field $K$ is a homomorphism $${\cal U}:G\longrightarrow GL(V)$$ of $G$ into the group $GL(V)$ of invertible linear transformations on $V$. The degree of $V$ is the degree of the representation $\cal U.$ If $G$ is a topological group and $\mathbb{U}({\cal H})$ is the group of unitary operators on the Hilbert space $\cal H,$ it is required that the homomorphism ${\cal U}:G\longrightarrow \mathbb U({\cal H})$ is strongly continuous, and differentiable in the case of $G$ a Lie group. We call $\cal U$ a unitary representation. A subspace $\mathcal{H}_0$ of $\mathcal{H}$ is said to be invariant under the unitary representation $\mathcal{U}$ if $\mathcal{U}_g\mathcal{H}_0\subset\mathcal{H}_0$ for all $g\in G$.  If the trivial subspace $\{0\}$ and $\mathcal{H}$ are the only invariant closed subspaces of $\mathcal{H}$ under $\mathcal{U}$, then $\mathcal{U}$ is irreducible.  It is the irreducible unitary representations that are the "atoms" of the unitary representations of $G$.

Two unitary representations of $G$, say $\mathcal{U}:G\rightarrow\mathbb{U}(\mathcal{H})$ and $\mathcal{U}':G\rightarrow\mathbb{U}(\mathcal{H}')$, are equivalent when there is an isometry  $A:\mathcal{H}\rightarrow\mathcal{H}'$ satisfying $A\circ\mathcal{U}_g=\mathcal{U}_g'\circ A$, for all $g\in G$.  So, the set of all unitary irreducible representations (UIRs) of $G$ can be partitioned into disjoint classes of UIRs.  A basic problem of representation theory of  Lie groups is the construction and classification of all UIRs, up to equivalence. In many cases the UIRs are sufficient to decompose $L^2-$functions on $G$ into their Fourier series or Fourier integral. In the compact group case, for example, the Peter-Weyl Theorem states that the matrix elements of the UIRs form a complete orthonormal set in $L^2(G).$ 

A good resource for a comprehensive list of representations of Lie groups is the 3-volume survey work of Vilenkin and Klimyk in \cite{Vilenkin2,Vilenkin3,Vilenkin4}.  For the Euclidean motion group $M(2)$, we recommend the earlier work of Vilenkin in \cite{Vilenkin} but in our discussion of its unitary representation, we compare ours with that of Sugiura in \cite[Chapter 4]{Sugiura}.

A more or less procedural way of constructing representations is  the method of induced representations by Frobenius and Mackey (for general groups, in \cite{Mackey3} and for locally compact groups, in \cite{Mackey2}). This is a method of  constructing representations of a group from representations of a subgroup. Let $\mathcal{S}$ be a representation of the subgroup $H$ on $V$ and $\mathcal{T}$ be the desired representation of $G$, induced by $\mathcal{S}$, that is $\mathcal{T}=\mathrm{Ind}^G_H$.  Let $L(G,H,V)$ be the space of functions $f:G\rightarrow V$ satisfying
	\begin{equation*}
		f(gh)=\mathcal{S}^{-1}_hf(g),
	\end{equation*}	 
for any $g\in G$ and $h\in H$.  Since $L(G,H,V)$ is invariant with respect to the left translation, the representation $\mathcal{T}$ of $G$ on $L(G,H,V)$ is defined by 
	\begin{equation*}
		(\mathcal{T}_gf)(g_0)=f(g^{-1}g_0).
	\end{equation*}

Specifically, we outline the construction of the representation of a semidirect product $G$, induced by its commutative subgroup $B$ (see \cite[Theorem 7.7]{Berndt1}).  Suppose $G=A\ltimes B$ and $A$ is a group of automorphisms on $B$.  The collection $X$ of 1-dimensional representations $\chi$ of $B$ is partitioned into disjoint orbits via the action $a\cdot\chi(b)=\chi(a^{-1}b)$ where $a\in A,b\in B$.  If $\Phi$ is one of these orbits, we define the collection of functions $\mathcal{H}=\{f:\Phi\rightarrow\mathcal{H}_0\}$ where $\mathcal{H}_0$ is the representation space of $\chi$.  Let $\phi\in\Phi$ and $\chi$ represents the class $\Phi$.  Then, the map $\mathcal{U}:G\rightarrow\mathrm{Aut}(\mathcal{H})$ defined by
	\begin{equation*}
		(\mathcal{U}_gf)(\phi)=\chi(b)f(a\cdot\phi)
	\end{equation*}	  
where $g=(a,b)$, is a representation of $G$, induced by the representation $\chi$ of $B$.

Let $M(2)$ be the Euclidean motion group of $2$ dimensions.  It is the semidirect product of $SO(2)$ and $\mathbb{R}^2$.  Its unitary representation \cite[p. 157]{Sugiura} is defined by
	\begin{equation}\label{UIR1}
		(\mathcal{U}^a_gf)(R_\theta)=e^{i(r,R_\theta a)}f(R^{-1}_\phi R_\theta)
	\end{equation}	   
where $g=(R_\phi,r)\in M(2)$, $f\in L^2(SO(2))$ and $a\in\mathbb{C}$.  This representation is induced by the $1$-dimensional unitary representation $\chi_a:r\mapsto e^{i(r,a)}$ of the commutative subgroup $\mathbb{R}^2$.  Since $\mathcal{U}^a$ is equivalent to $\mathcal{U}^b$ if and only if $|a|=|b|$ \cite[Chapter IV Theorem 1.3]{Sugiura}, an equivalence class of UIRs of $M(2)$ can be represented by $\mathcal{U}^a$ where $a>0$.   Since $SO(2)\simeq S^1\ni(\cos\theta,\sin\theta)$, letting $r=(r_1,r_2)$, expression (\ref{UIR1}) becomes
	\begin{equation}\label{UIR2}
		(\mathcal{U}^a_gf)(\theta)=e^{ia(r_1\cos\theta+r_2\sin\theta)}f(\theta-\phi).
	\end{equation}
The set  $P=\{\mathcal{U}^a:a>0\}$ of infinite-dimensional UIRs is called the principal series of UIRs of $M(2)$.

There is another set of UIRs other than the set $P$.   These representations are the $1$-dimensional unitary representations $\chi_n,n\in\mathbb{Z}$ of $SO(2)$ via the natural projection $p:M(2)\rightarrow SO(2)$,   defining the operators  
	\begin{equation}\label{UIR1dim}
		(\chi_n\circ p)(R_\phi,r)=e^{in\phi}.
	\end{equation}
Hence, the complete set of representatives of the set of classes of UIRs of $M(2)$ \cite[Chapter IV Theorem 2.1]{Sugiura} is
	\begin{equation}\label{DualofM(2)}
		\widehat{M(2)}=\{\mathcal{U}^a:a>0\}\cup\{\chi_n\circ p:n\in\mathbb{Z}\}.
	\end{equation}	  

At this point, consider the infinite-dimensional UIR $\mathcal{U}^a$.  Let $U$ be an element of the Lie algebra $\mathfrak{m}(2)=\mathrm{span}\{X,E_1,E_2\}$ of the Euclidean motion group $M(2)$ where $X$ spans the Lie algebra of $SO(2)$, $E_1,E_2$ are the canonical base elements that span $\mathbb{R}^2$ and the Lie brackets of these spanning elements are $[X,E_1]=-E_2,[X,E_2]=E_1$ and $[E_1,E_2]=0$.  Given by the $1$-parameter subgroup 
	\begin{equation*}
		\exp tU=\left\{ \begin{array}{l l}
			\bigg(R_{-tc_1},\bigg(\frac{c_2}{c_1}\sin tc_1+\frac{c_3}{c_1}(1-\cos tc_1),\\
			\frac{c_2}{c_1}(-1+\cos tc_1)+\frac{c_3}{c_1}\sin tc_1\bigg)\bigg) & \text{if }c_1\not=0 \\
			& \\
			(1,(tc_2,tc_3)) & \text{if }c_1=0
		\end{array} \right.
	\end{equation*}
of $M(2)$ where $U=c_1X+c_2E_1+c_3E_2$, expression (\ref{UIR2}) becomes
	\begin{equation}\label{UIR1Para}
		(\mathcal{U}^a_{\exp tU}f)(\theta)=\left\{ \begin{array}{l l}
			e^{ia\left[\frac{c_2}{c_1}(\sin(tc_1+\theta)-\sin\theta)-\frac{c_3}{c_1}(\cos(tc_1+\theta)-\cos\theta)\right]}&\\
			\times f(tc_1+\theta) & \text{if }c_1\not=0 \\
			& \\
			e^{iat(c_2\cos\theta+c_3\sin\theta)}f(\theta) & \text{if }c_1=0
		\end{array} \right. 
	\end{equation}
and its derivative with respect to $t$ is
	\begin{equation}\label{UIRDer}
		\frac{d}{dt}\mathcal{U}^a_{\exp tU}f(\theta)=\left\{\begin{array}{l l}
		e^{ia\left[\frac{c_2}{c_1}(\sin(tc_1+\theta)-\sin\theta)-\frac{c_3}{c_1}(\cos(tc_1+\theta)-\cos\theta)\right]} &  \\
		\times\big[ia(c_2\cos(tc_1+\theta)+c_3\sin(tc_1+\theta)) & \\
		\times f(tc_1+\theta) & \text{if }c_1\not= 0 \\
		+c_1\frac{\partial}{\partial(tc_1+\theta)}f(tc_1+\theta)\big] & \\
			& \\
		ia(c_2\cos\theta+c_3\sin\theta)(\mathcal{U}^a_{\exp tU}f)(\theta) & \text{if }c_1=0 
\end{array}	\right.		 
	\end{equation}
and when $t=0$
	\begin{equation}\label{UIRDer1}
		(d\mathcal{U}^a(U)f)(\theta)=ia(c_2\cos\theta+c_3\sin\theta)f(\theta)+c_1f'(\theta),
	\end{equation}	
where $d\mathcal{U}^a(U)=\frac{d}{dt}\mathcal{U}^a_{\exp tU}|_{t=0}$.
\section{Quantization}\label{Quantization}

Quantization is the process of forming a quantum mechanical system from a given classical system where these two systems, classical mechanics (in the Hamiltonian formalism) and quantum mechanics (in the Heisenberg picture), are modeled by the space of $C^\infty$-functions on a symplectic manifold $M$ and the set of self-adjoint operators on a Hilbert space $\mathcal{H}$, respectively.  This is done by associating a classical observable $f$ on $M$ to a self-adjoint operator $Q(f)$ on $\mathcal{H},$ where $Q$ is a linear map, $Q(1)$ is the identity operator and satisfies the correspondence
	\begin{equation*}
		Q(\{f,g\})=-\frac{i}{\hbar}[Q(f),Q(g)]
	\end{equation*}
where the expression above was the result of Dirac's analogy of Heisenberg commutator bracket $[\cdot,\cdot]$ with the Poisson bracket $\{\cdot,\cdot\}$, which endow the two respective mechanical systems their Lie algebra structures.  

When $M=T^*N$, where $N$ is an $n$-dimensional smooth manifold and $\mathcal{H}=L^2(N)$, the quantization is said to be full if the operators $Q(q^i)$ and $Q(p_j)$ act irreducibly on $\mathcal{H}$.  That is, the operators above are the position and momentum operators: $Q(q^i)$ is the multiplication of $q^i$ and $Q(p_j)=-i\hbar\partial_{q_j}$. By the theorem of Stone and von Neumann, it is unitarily equivalent to the Schr\"{o}dinger representation.  

It is known that the algebra of inhomogenous quadratic polynomials on $\mathbb{R}^{2n}$ is a maximal Lie subalgebra of the space of polynomials under the Poisson bracket.  This subalgebra is identified with the Lie algebra of the Jacobi group.  A representation of this group, known as the Schr\"{o}dinger-Weil representation, gives rise to a quantization map.  However, by the Groenewold-van Hove theorem, it is impossible to extend this map to the whole $C^\infty(\mathbb{R}^{2n})$ (see \cite[Chapter V]{Berndt}).  

Independently, the geometric quantization of Konstant and Souriau is another Hilbert space-based quantization where the goal is the construction of quantum objects from the geometry of the classical ones \cite[p. 138]{Kirillov}.  This quantization procedure is the physical counterpart of Kirillov's orbit method. An orbit of a Lie group $G$ in the coadjoint representation, also known as coadjoint orbit, is the orbit of the coadjoint action of $G$ on the dual $\mathfrak{g}^*$ of its Lie algebra $\mathfrak{g}$, through the point $F\in\mathfrak{g}^*$.  It is given by the set
	\begin{equation}\label{CoadjointOrbit}
		\Omega=\{K(g)F:g\in G\}
	\end{equation}	 
where
	\begin{equation*}
		\left<K(g)F,U\right>=\left<F,\mathrm{Ad}_{g^{-1}}U\right>
	\end{equation*}
and $\left<\cdot,\cdot\right>$ is the dual pairing of the Lie algebra with its dual.  It is known that the coadjoint orbit $\Omega$ is a homogeneous symplectic $G$-manifold \cite[Theorem 1.1]{Kirillov} and its symplectic form $\omega$ is called the Kirillov symplectic form.

This method's particular interest is the correspondence between the finite-dimensional coadjoint orbits and the infinite-dimensional unitary representations of $G$.  The method first appeared in its application to nilpotent Lie groups \cite{Kirillov1} and further extended to other classes of Lie groups (see \cite{Kirillov2} and \cite{Kirillov3}).

In both of the methods above, classical mechanics is a limiting case (that is, $\hbar\rightarrow 0$ in Dirac's correspondence principle) of quantum mechanics \cite{Dirac}. Moreover, in these definitions of quantization, the association of a $C^\infty$-function to a self-adjoint operator is quite a radical transition. In the next section, we define a quantization method free from the Hilbert space-based formulation of quantum mechanics.

\section{Deformation Quantization}\label{DeformationQuantization}

Earlier, we have briefly introduced deformation quantization or phase-space quantum mechanics. The model of quantum mechanics is described as a deformed structure of the space of classical observables.  In this deformed structure, a noncommutative but associative product is introduced, called the $\star$-product.  

Let $f,g\in C^\infty(M)$ where $M$ is a Poisson manifold.  This formal associative $\star$-product \cite{Gutt}, here we denote this as $\star_\lambda$, is a bilinear map
	\begin{equation*}
		C^\infty(M)\times C^\infty(M)\rightarrow C^\infty(M)[[\lambda]]
	\end{equation*}
defined by
	\begin{equation*}
		f\star_\lambda g=\sum_{r=0}^\infty \lambda^r C^r(f,g)
	\end{equation*}
where $\lambda$ is a formal parameter, $C^r$ is a bidifferential operator with $C^r(f,g)=(-1)^rC^r(g,f)$ for all $f,g\in C^\infty(M)$ and satisfies the following properties:
	\begin{enumerate}
		\item[1.] $C^0(f,g)=fg$
		\item[2.] $C^1(f,g)=\{f,g\}$ and
		\item[3.] $C^r(1,f)=C^r(f,1)=0$ for $r\geq 1$.
	\end{enumerate}			  
Property 1 shows that the noncommutative product $\star_\lambda$ is a deformation of the commutative pointwise multiplication of functions in $C^\infty(M)$.  Property 2 satisfies the correspondence principle
	\begin{equation*}
		f\star_\lambda g-g\star_\lambda f=2\lambda\{f,g\}+\cdots		
	\end{equation*}
where the dots mean higher-order terms with respect to $\lambda$ and if we let
	\begin{equation*}
		[f,g]_\lambda=\frac{1}{2\lambda}(f\star_\lambda g-g\star_\lambda f),
	\end{equation*}
the bracket $[\cdot,\cdot]_\lambda$ is the deformed Poisson bracket in $C^\infty(M)$.  Property 3 implies $1\star_\lambda f=f\star_\lambda 1=f$.  Hence, the algebra $(C^\infty(M)[[\lambda]],\star_\lambda,[\cdot,\cdot]_\lambda)$ is the quantum analogue of the classical model $(C^\infty(M),\cdot,\{\cdot,\cdot\})$.  The questions of existence and classification of these $\star$-products have already been settled (see the review in \cite{Bordemann}).

The $\star$-product for the symplectic flat  manifold $M=\mathbb{R}^{2n}$ has long been known \cite{Moyal,Groenewold} and is the most important. We discuss it at length.  Suppose $\omega$ is the canonical symplectic form of $M$ in the $(q,p)$ coordinates on some open set $O\subset M$, the Moyal $\star$-product of the algebra $(C^\infty(M)[[\lambda]],\star)$ with $\lambda=\frac{1}{2i}$ is the product
	\begin{equation}\label{Moyal}
		f\star g=fg+\sum_{r=1}^\infty \frac{1}{r!}\left(\frac{1}{2i}\right)^rP^r(f,g)
	\end{equation}	  
where 
	\begin{equation*}
		P^r(f,g)=\Lambda^{i_1j_1}\Lambda^{i_2j_2}\cdots\Lambda^{i_rj_r}\partial_{i_1i_2\cdots i_r}f\partial_{j_1j_2\cdots j_r}g
	\end{equation*}
with the multi-index notation
	\begin{equation*}
		\partial_{i_1i_2\cdots i_r}=\frac{\partial}{\partial x_{i_1}\partial x_{i_2}\cdots\partial x_{i_r}},x:=(q^1,...,q^n,p_1,...,p_n)
	\end{equation*}
and $\Lambda^{ij}$ are constant value entries of the matrix associated to the symplectic form $\omega$. 

This $\star$-product has an integral formula \cite{Hansen}, from which many of its important properties follow directly.  Let $f,g$ be functions in the Schwartz space $\mathcal{S}(\mathbb{R}^{2n})$.  By defining the symplectic Fourier transform $F:\mathcal{S}(\mathbb{R}^{2n})\rightarrow\mathcal{S}(\mathbb{R}^{2n})$ by
	\begin{equation*}
		(Ff)(x)=\int_{\mathbb{R}^{2n}}f(\xi)e^{i\omega(x,\xi)}\frac{d\xi}{(2\pi)^n}
	\end{equation*}
and the symplectic convolution $\times_\omega$ as
	\begin{equation*}
		(f\times_\omega g)(x)=\int_{\mathbb{R}^{2n}}f(t)g(x-t)e^{i\omega(t,x)}\frac{dt}{(2\pi)^n},
	\end{equation*}
the product
	\begin{equation*}
		f\star g=F(Ff\times_\omega Fg)
	\end{equation*}
admits the development of the Moyal $\star$-product defined in (\ref{Moyal}) which converge to a function in $\mathcal{S}(\mathbb{R}^{2n})$ and has the following properties:
	\begin{enumerate}
		\item[1.] $(\mathcal{S}(\mathbb{R}^{2n}),\star)$ is a generalized Hilbert algebra in $L^2(\mathbb{R}^{2n})$;
		\item[2.] $\int (f\star g)(\xi)d\xi=\int (fg)(\xi) d\xi$;
		\item[3.] $\overline{f\star g}=\bar{g}\star\bar{f}$; and 
		\item[4.] the operator $l_f:\mathcal{S}(\mathbb{R}^{2n})\rightarrow\mathcal{S}(\mathbb{R}^{2n})$ defined by $l_f(g)=f\star g$, can be extended to a bounded operator on $L^2(\mathbb{R}^{2n})$.
	\end{enumerate}

Bayen et al. \cite[p. 132]{Bayen1} predicted that deformation quantization has a promising future in representation theory.  Motivated by Kirillov's orbit method via Konstant and Souriau's geometric quantization and Fronsdal's initial investigation in \cite{Fronsdal}, D. Arnal together with J.C. Cortet, J. Ludwig, M. Cahen and S. Gutt, wrote a series of articles about the application of deformation theory on representations of general classes of Lie groups: nilpotent Lie groups \cite{Arnal1,Arnal2}, compact Lie groups \cite{Arnal4}, exponential Lie groups \cite{Arnal3}, and solvable Lie groups \cite{Arnal7}.  These computations were made possible due to the covariance property of the Moyal $\star$-product \cite{Arnal5}.    

For a unitary representation of a connected Lie group $G$ corresponding to an orbit $\Omega\simeq G/G_F$, where $G_F$ is the stabilizer subgroup of $G$, the Lie algebra $\mathfrak{g}$ is identified with the Lie subalgebra of $C^\infty(\Omega)$
	\begin{equation*}
		\mathfrak{g}_\Omega=\{\tilde{U}\in C^\infty(\Omega):U\in\mathfrak{g}\}
	\end{equation*}
where the function $\tilde{U}:\Omega\rightarrow\mathbb{R}$ is defined by 
	\begin{equation}\label{Hamiltonian}
		\tilde{U}(F)=\left<F,U\right>
	\end{equation}
for all $F\in\Omega$ and one has to show that the Moyal $\star$-product satisfies
	\begin{equation}\label{RelativeQuantization}
		\frac{1}{2\lambda}(\tilde{U}\star\tilde{T}-\tilde{T}\star\tilde{U})=\widetilde{[U,T]}
	\end{equation}
for any $U,T\in\mathfrak{g}$.  A $\star$-product that satisfies expression (\ref{RelativeQuantization}) is a $\mathfrak{g}_\Omega$-relative quantization.  The main result of the paper \cite{Arnal5} is that each quantization relative to a Lie algebra $\mathfrak{g}$ is a $G$-covariant $\star$-product, and a $G$-covariant $\star$-product gives rise to a representation $\tau$ of $G$ on $C^\infty(\Omega)[[\lambda]]$ by automorphisms, which also gives rise to a differential representation $d\tau$ of $\tau$, defined by $d\tau(U)=\left.\frac{d}{dt}\tau(\exp tU)\right|_{t=0}$. That is, we obtain a representation of $\mathfrak{g}$ on $C^\infty(\Omega)[[\lambda]]$ by endomorphisms.

The function $\tilde{U}$ on $\Omega$ is called the Hamiltonian function associated to the Hamiltonian vector field $\xi_U$, defined by $\xi_Uf=\{\tilde{U},f\}$. We remark that the computations above depend on the parameterization of the orbit $\Omega$.

The computational techniques that were outlined in the construction of representations of nilpotent \cite{Arnal1,Arnal2} and exponential \cite{Arnal3} Lie groups have led to concrete computations of representations for particular Lie groups, some of which were neither nilpotent nor exponential.  Among these are the works of Diep and his students: the group of affine transformation of the real and complex plane \cite{Diep1,Diep2}, the real rotation groups \cite{Nable} and the $MD_4$-groups \cite{Nguyen}.    The orbits generated by the group of affine transformation of the complex plane and the real rotation groups were parameterized by local charts, while the others have global charts.

These papers have provided us an outline to construct and classify unitary representations of concrete Lie groups.  As in the method of obtaining representations via induction, we have a more or less procedural way of the construction.  Our main contribution is the development of the UIRs of $M(2)$ via deformation quantization, hence an alternative to the method of induced representation.  The construction in the next section is outlined as follows:
	\begin{enumerate}
		\item[1.] compute the coadjoint orbit $\Omega_F$ of $M(2)$ through the point $F\in\mathfrak{m}(2)^*$;
		\item[2.] define a chart on $\Omega_F$ and consider the Hamiltonian system $(\Omega_F,\omega,\xi_U)$ where the Hamiltonian function $\tilde{U}$ is defined in (\ref{Hamiltonian}), $\xi_U$ is its associated vector field and $\omega$ is the Kirillov symplectic form;
		\item[3.] the Moyal $\star$-product is $M(2)$-covariant which will give rise to a representation $l$ of $\mathfrak{m}(2)$ on $C^\infty(\Omega_F)[[\lambda]]$;
		\item[4.] the representation $\hat{l}$, defined by the operators $\hat{l}_U=\mathcal{F}_p\circ l_U\circ\mathcal{F}^{-1}_p$, is a differential representation of the UIR of $M(2)$ where the operator $\mathcal{F}_p$ is a partial Fourier transform; and
		\item[5.] classify these constructed representations via the coadjoint orbits.
	\end{enumerate}	     
We remark that these steps are quite straightforward to implement and provide concrete computations suitable for the learning by graduate students in Physics and Mathematics of many important mathematical concepts and objects.  

\section{The UIRs of \boldmath$M(2)$}\label{UIRM(2)}

\subsection{Coadjoint orbits}
In matrix form, the Lie algebra $\mathfrak{m}(2)$ of $M(2)$ is spanned by the matrices
	 	\begin{equation*}
		X=\left(\begin{array}{c c c}
			0 & 1 & 0 \\
			-1& 0 & 0 \\
			0 & 0 & 0
		\end{array}\right),
		E_1=\left(\begin{array}{c c c}
			0 & 0 & 1 \\
			0 & 0 & 0 \\
			0 & 0 & 0
		\end{array}\right),
		E_2=\left(\begin{array}{c c c}
			0 & 0 & 0 \\
			0 & 0 & 1 \\
			0 & 0 & 0
		\end{array}\right)
	\end{equation*}
and these matrices satisfy the Lie brackets $[X,E_1]=-E_2,[X,E_2]=E_1$ and $[E_1,E_2]=0$.  Hence, $\mathfrak{m}(2)$ is identified with $\mathbb{R}\times\mathbb{R}^2$ and the elements are written as $U=c_1X+c_2E_1+c_3E_2$.  The dual $\mathfrak{m}(2)^*$ is also identified with $\mathbb{R}\times\mathbb{R}^2$.	  

Let $g=\exp U\in M(2)$ and fix $F=(\mu,\alpha)=\mu X^*+\alpha_1E_1^*+\alpha_2E_2^*\in\mathfrak{m}(2)^*$.  The coadjoint orbit  $\Omega_F$ of $M(2)$ through $F$, given by expression (\ref{CoadjointOrbit}), is the set
	\begin{equation*}
		\Omega_F=\{K(\exp U)F: U\in\mathfrak{m}(2)\}\subset\mathfrak{m}(2)^*
	\end{equation*}
satisfying
	\begin{equation*}
		\left<K(\exp U)F,T\right>=\left<F,\mathrm{Ad}(-\exp U)T\right>.
	\end{equation*}
We write
	\begin{eqnarray*}
		K(\exp U)F&=&\left<F,\exp(-\mathrm{ad}_U)X\right>X^*+\left<F,\exp(-\mathrm{ad}_U)E_1\right>E_1^* +\left<F,\exp(-\mathrm{ad}_U)E_2\right>E_2^*.
	\end{eqnarray*}
    
But
	\begin{equation*}
		\exp(-\mathrm{ad}_U)=\sum_{r\geq 0}\frac{1}{r!}\left(\begin{array}{c c c}
			 0 & 0 & 0\\
			 c_3 & 0 &-c_1\\
			-c_2 & c_1 & 0
		\end{array} \right)^r=
		\left(\begin{array}{c c}
			1 & 0 \\
			\frac{1-R_{c_1}}{c_1}\left(\begin{array}{c}
			c_2\\
			c_3
			\end{array}\right) & R_{c_1}
		\end{array}\right).
	\end{equation*}
So we have
	\begin{equation*}
		K(\exp U)F=\left(\mu+\alpha\cdot\frac{1-R_{c_1}}{c_1}\left(\begin{array}{c}
			c_2\\
			c_3
		\end{array}\right)\right)X^*+\alpha R_{c_1}\left(\begin{array}{c}
			E_1^*\\
			E_2^*
		\end{array}\right).
	\end{equation*}
The coadjoint orbit of $M(2)$ through $F$ is
	\begin{equation*}
		\Omega_F=\left\{\left(\mu+\alpha\cdot\frac{1-R_{c_1}}{c_1}\left(\begin{array}{c}
			c_2\\
			c_3
		\end{array}\right),\alpha R_{c_1}\right): U\in\mathfrak{m}(2)\right\}.
	\end{equation*}

There are two types of orbits.  If $\alpha=0$, the orbit $\Omega_F=\{(\mu,0)\})$ is a point- the trivial orbit.  If $\alpha\not=0$, the orbit $\Omega_F$ is the 2-dimensional infinite cylinder of radius $\|\alpha\|$ which we denote $\Omega_F=T^*S^1_{\|\alpha\|}$.     We first work on the nontrivial orbits, then later the trivial ones.

\subsection{Hamiltonian system on the cylinder}

Fix $F$ where $\alpha\not=0$.  The map
	\begin{equation}\label{Chart}
		\psi:\mathbb{R}^2\rightarrow \Omega_F=T^*S^1_{\|\alpha\|}
	\end{equation}
where $\psi(x,\theta)=xX^*+\|\alpha\|\cos\theta E_1^*+\|\alpha\|\sin\theta E_2^*$ defines a global chart on $\Omega_F$. So each function $f$ in $C^\infty(\Omega_F)$ is written as $f\circ\psi$ and we describe the Hamiltonian system $(\Omega_F,\omega,\xi_U)$ with respect to the chart (\ref{Chart}) as follows:

	\begin{enumerate}
		\item[1.] the Hamiltonian function associated to $U\in\mathfrak{m}(2)$ is 
	\begin{equation}\label{HamiltonianFunc}
		\tilde{U}=c_1x+\|\alpha\|(c_2+ic_3,e^{i\theta})
	\end{equation}			 
where $(\cdot,\cdot)$ is  the inner product and the associated Hamiltonian vector field is 
	\begin{equation*}
		\xi_U=c_1\partial_\theta-\|\alpha\|(c_2+ic_3,ie^{i\theta})\partial_x;
	\end{equation*}
		\item[2.] the map $\psi$ gives rise to a symplectomorphism where the Kirillov symplectic form is the canonical form $\omega=dx\wedge d\theta$. 
	\end{enumerate}

Since $U=c_1X+c_2E_1+c_3E_2\in\mathfrak{m}(2)$, the value of the functional $\tilde{U}$ at the point $F'=xX^*+\|\alpha\|\cos\theta E_1^*+\|\alpha\|\sin\theta E_2^*\in\Omega_F$ is the value of the dual pairing
		\begin{equation*}
			\left<F',U\right>=c_1x+c_2(\|\alpha\|\cos\theta)+c_3(\|\alpha\|\sin\theta),
		\end{equation*}
and since $\xi_Uf=\partial_x\tilde{U}\partial_\theta f-\partial_\theta\tilde{U}\partial_x f$ in $(x,\theta)$-coordinates, it follows that
		\begin{equation*}
			\xi_U=c_1\partial_\theta-\|\alpha\|(-c_2\sin\theta+c_3\cos\theta)\partial_x.
		\end{equation*}
The restriction of $\psi$ to the domain $\mathbb{R}\times\mathbb{T}$ gives rise to a diffeomorphism.  Let $U=c_1X+c_2E_1+c_3E_2$ and $T=c'_1X+c'_2E_1+c'_3E_2$. Since $[U,T]=(c_1c'_3-c'_1c_3)E_1+(c'_1c_2-c_1c'_2)E_2$, so for any $F'\in\Omega_F$
		\begin{equation}\label{bracketUT}
			\left<F',[U,T]\right>=\|\alpha\|\cos\theta(c_1c'_3-c'_1c_3)+\|\alpha\|\sin\theta(c'_1c_2-c_1c'_2).
		\end{equation}
But
		\begin{eqnarray*}
			\omega(\xi_U,\xi_T)&=&\mathrm{det}\left(\begin{array}{c c}
			dx(\xi_U)&dx(\xi_T) \\
		d\theta(\xi_U) & d\theta(\xi_T)	
			\end{array}\right) \\
			&=&\|\alpha\|\cos\theta(c_1c'_3-c'_1c_3)+\|\alpha\|\sin\theta(c'_1c_2-c_1c'_2),
		\end{eqnarray*}
when $\omega=dx\wedge d\theta$.  Hence, $\psi|_{\mathbb{R}\times\mathbb{T}}$ is a symplectomorphism.

\subsection{Covariance of the Moyal $\star-$product}

Let $\Lambda$ be the matrix associated to the canonical form $\omega=dx\wedge d\theta$, that is, 
	\begin{equation*}
		\Lambda=\left(\begin{array}{c c}
			0 & 1 \\
			-1 &0
		\end{array} \right).
	\end{equation*}
The Moyal $\star$-product is defined by expression (\ref{Moyal}) where $\lambda=\frac{1}{2i}$.  Since $P^0(\tilde{U},\tilde{T})=\tilde{U}\tilde{T}$, $P^1(f,g)=\partial_x\tilde{U}\partial_\theta\tilde{T}-\partial_\theta\tilde{U}\partial_x\tilde{T}=\|\alpha\|\cos\theta(c_1c'_3-c'_1c_3)+\|\alpha\|\sin\theta(c'_1c_2-c_1c'_2)$ and $P^r(\tilde{U},\tilde{T})=0$ for $r\geq 2$, we have 
	\begin{equation}\label{UstarT}
		\tilde{U}\star\tilde{T}=\tilde{U}\tilde{T}+\frac{1}{2i}(\|\alpha\|\cos\theta(c_1c'_3-c'_1c_3)+\|\alpha\|\sin\theta(c'_1c_2-c_1c'_2)).
	\end{equation}
So from (\ref{UstarT}), we can easily compute
	\begin{eqnarray}\label{iUstariT}
		i\tilde{U}\star i\tilde{T}-i\tilde{T}\star i\tilde{U}=i\widetilde{[U,T]}
	\end{eqnarray}
where $\widetilde{[U,T]}$ is expression (\ref{bracketUT}).

Expression (\ref{RelativeQuantization}) is exactly (\ref{iUstariT}) when $\lambda=\frac{1}{2i}$.  Thus, the Moyal $\star$-product is $M(2)$-covariant.  Hence, it gives rise to a representation of $\mathfrak{m}(2)$ on $C^\infty(\Omega_F)[[\lambda]]$ by endomorphism of the Moyal $\star$-product. 

This representation of $\mathfrak{m}(2)$ is defined by the operators 
	\begin{equation*}
		l_U:C^\infty(\Omega_F)[[\lambda]]\rightarrow C^\infty(\Omega_F)[[\lambda]]
	\end{equation*}
given by the left $\star$-product multiplication
	\begin{equation*}
		l_Uf=\frac{1}{2\lambda}\tilde{U}\star f.
	\end{equation*}
But as we have earlier explained, the Moyal $\star$-product converges in the space $\mathcal{S}(\Omega_F)$ and that the operator $l_U$ extends to $L^2(\Omega_F)$. We still denote this extension as $l_U$ for all $U\in\mathfrak{m}(2)$.	
	
\subsection{Convergence of the operators $\hat{l}_U$}	

Instead of $l_U$, we will compute for the convergence of $\hat{l}_U=\mathcal{F}_x\circ l_U\circ\mathcal{F}^{-1}_x$, for all $U\in\mathfrak{m}(2)$ as suggested in \cite{Arnal2,Arnal3}.  In the case of exponential Lie groups, $\hat{l}$ is the differential of the UIR of the said group associated to the orbit method of Kostant-Kirillov \cite[Proposition 2.6]{Arnal3}.  Both the exponential Lie groups and $M(2)$ are solvable, but the latter is non-exponential (since the exponential map $\exp:\mathfrak{m}(2)\rightarrow M(2)$ is not injective).  However, we will show in section \ref{RefNontrivial} that $\hat{l}$ is the differential of the UIR of $M(2)$.     	
	
Let $f\in\mathcal{S}(\Omega_F)$.  The partial Fourier transform $\mathcal{F}_x$ of the function $f$ on $\Omega_F$ is defined by
	\begin{equation*}
		(\mathcal{F}_xf)(\eta,\theta)=\int_\mathbb{R}e^{-i\eta x}f(x,\theta)\frac{dx}{\sqrt{2\pi}}	
	\end{equation*}	
and its inverse transform $\mathcal{F}^{-1}_x$ by
	\begin{equation*}
		(\mathcal{F}^{-1}_xf)(x,\theta)=\int_\mathbb{R}e^{i\eta x}f(\eta,\theta)\frac{d\eta}{\sqrt{2\pi}}.	
	\end{equation*}	

The derivatives
\begin{equation}\label{dPFT}
\partial_x\mathcal{F}^{-1}_x(f)=i\mathcal{F}^{-1}_x(\eta f)
\end{equation}
and 
\begin{equation}\label{PFTd}
\mathcal{F}_x(xf)=i \partial_\eta \mathcal{F}_x(f)
\end{equation}
are easily computed while the derivative $(\ref{dPFT})$ can be generalized as
\begin{equation}\label{drPFT}
	\frac{\partial^r}{\partial{x^r}}\mathcal{F}^{-1}_x(f)=i^r\mathcal{F}^{-1}_x(\eta^rf)
	\end{equation}	

On the other hand, the partial derivative of $\tilde{U}$ with respect to $x$ of order $r\geq 2$ or with respect to a mixture of variables $x$ and $\theta$ is zero. So, the bidifferential $P^r(\tilde{U},\mathcal{F}^{-1}_xf)$ will always have the nonzero term
	\begin{equation}\label{bdiff}
		\Lambda^{21}\Lambda^{21}\cdots\Lambda^{21}\partial_{\theta^r}\tilde{U}\partial_{x^r}\mathcal{F}^{-1}_x(f)
	\end{equation}
where $\Lambda^{21}\Lambda^{21}\cdots\Lambda^{21}$ $r$-times.  The $r$th partial derivative of expression (\ref{HamiltonianFunc}), together with generalized derivative (\ref{drPFT}) applied in (\ref{bdiff}), we have  
	\begin{equation}\label{diffPr}
	P^r(\tilde{U},\mathcal{F}^{-1}_xf)=(-1)^r\|\alpha\|(c_2+ic_3,i^re^{i\theta})(i^r\mathcal{F}^{-1}_x)(\eta^rf) 
	\end{equation}	
for $r \geq 2$ for all functions $f$ on $\Omega_F$.

Now $\hat{l}_U(f)=i\mathcal{F}_x(\tilde{U} \star \mathcal{F}^{-1}_x(f))$.  Applying (\ref{dPFT}) and (\ref{diffPr}), we have
	\begin{eqnarray*}
		\tilde{U} \star \mathcal{F}^{-1}_x(f)&=& c_1x \mathcal{F}^{-1}_x(f)+\frac{c_1}{2i} \partial_\theta \mathcal{F}^{-1}_x(f) \\
		&&+\sum_{r=0}^\infty \frac{1}{r!}\left(-\frac{1}{2}\right)^r \|\alpha \|(c_2+ic_3,i^re^{i\theta})\mathcal{F}^{-1}_x(\eta^r \cdot f).
	\end{eqnarray*}
Together with (\ref{PFTd}),
	\begin{eqnarray*}
		\hat{l}_U(f)&=&-c_1\partial_\eta f+\frac{c_1}{2} \partial_\theta f \\
		&&+ i\|\alpha\|\sum_{r=0}^\infty \frac{1}{r!}\left(-\frac{\eta}{2}\right)^r (c_2+ic_3,i^re^{i\theta})f \\
		&=&c_1\left(\frac{1}{2} \partial_\theta-\partial_\eta \right)f+i\|\alpha\|\left(c_2+ic_3,e^{i\theta} \sum_{r=0}^\infty \frac{1}{r!}\left(-\frac{i\eta}{2}\right)^r \right)f \\
		&=&c_1\left(\frac{1}{2} \partial_\theta-\partial_\eta \right)f+i\|\alpha\|\left(c_2+ic_3,e^{i\left(\theta-\frac{\eta}{2}\right)} \right)f
	\end{eqnarray*}
Let $s=\theta-\frac{\eta}{2}$.  By the change of variables, the above expression will become 
\begin{equation}\label{DifferentialOperatorsM(2)Corollary}
		\hat{l}_U=c_1\frac{\partial}{\partial s}+i \|\alpha \|(c_2\cos s+c_3 \sin s).
	\end{equation}		

\subsection{Representations associated to the nontrivial orbits}\label{RefNontrivial}

The representation $\hat{l}$ of $\mathfrak{m}(2)$ on $L^2(\Omega_F)$ is defined by the operators $\hat{l}_U$ in (\ref{DifferentialOperatorsM(2)Corollary}).  But the representation space $L^2(\Omega_F)$ is too big.  We choose $\mathfrak{h}=\mathbb{R}^2$ as the real algebraic polarization of $\mathfrak{m}(2)$.  By Remark 6 in \cite[p. 29]{Kirillov2}, the leaves of the $M(2)-$invariant foliation of $\Omega_F$ are the disjoint tangent lines passing through each point in $S^1_{\|\alpha\|}$.  This means that the subalgebra of functions on $\Omega_F$ which are constant along these leaves is a maximal abelian subalgebra of $C^\infty(\Omega_F)$.  Hence, we reduce $L^2(\Omega_F)$ into $L^2\left(S^1_{\|\alpha\|}\right)$.  Furthermore, $L^2\left(S^1_{\|\alpha\|}\right)$ is isomorphic to $L^2(S^1)$ given by the map $f\mapsto f|_{S^1}$, where $(s_1,s_2)\in S^1_{\|\alpha\|}$ is identified with $\left(\frac{s_1}{\|\alpha\|},\frac{s_2}{\|\alpha\|}\right)\in S^1$.  So, $\hat{l}$ is a representation of $\mathfrak{m}(2)$ in $L^2(S^1)$.  We are left to show that $\hat{l}$ is the differential of the unitary representation of $M(2)$ defined in (\ref{UIR2}).

Set $\|\alpha\|=a$ and $s=\theta$ in (\ref{DifferentialOperatorsM(2)Corollary}).  But this is exactly (\ref{UIRDer1}).  To show uniqueness, we apply the differential operator $\hat{l}_U$ to expression (\ref{UIR1Para}) where $c_1\not=0$, so
	\begin{eqnarray}\label{lU}
		\hat{l}_U(\mathcal{U}^a_{\exp tU}f)(\theta)&=&ia(c_2\cos\theta+c_3\sin\theta)(\mathcal{U}^a_{\exp tU}f)(\theta)+c_1\frac{\partial}{\partial\theta}(\mathcal{U}^a_{\exp tU}f)(\theta).
	\end{eqnarray}
The second term in (\ref{lU}) is computed as
	\begin{eqnarray}\label{lU1}
		c_1\frac{\partial}{\partial\theta}(\mathcal{U}^a_{\exp tU}f)(\theta)&=&e^{ia[\frac{c_2}{c_1}(\sin(tc_1+\theta)-\sin\theta)-\frac{c_3}{c_1}(\cos(tc_1+\theta)-\cos\theta)]}\nonumber\\
		&&\bigg(ia[c_2(\cos(tc_1+\theta)-\cos\theta) \nonumber \\
		&&+c_3(\sin(tc_1+\theta)-\sin\theta)]f(tc_1+\theta)\nonumber\\
		&&+c_1f'(tc_1+\theta)\bigg).
	\end{eqnarray}
When (\ref{lU1}) replaces the second term in (\ref{lU}),
	\begin{equation*}
		\frac{d}{dt}(\mathcal{U}^a_{\exp tU}f)(\theta)=\hat{l}_U(\mathcal{U}^a_{\exp tU}f)(\theta)
	\end{equation*}
where the left-hand side is the derivative of $\mathcal{U}^a$ expressed in (\ref{UIRDer}).

Consider $c_1=0$.  The left-hand side in (\ref{UIRDer}) together with the application of the operator $\hat{l}_U=ia(c_2\cos\theta+c_3\sin\theta)$ to the expression (\ref{UIR1Para}), will result to the equality
	\begin{eqnarray*}
		\frac{d}{dt}(\mathcal{U}^a_{\exp tU}f)(\theta)&=&ia(c_2\cos\theta+c_3\sin\theta)(\mathcal{U}^a_{\exp tU}f)(\theta)\nonumber\\
		&=&\hat{l}_U(\mathcal{U}^a_{\exp tU}f)(\theta).
	\end{eqnarray*}
	
For both cases, the derivative with respect to  $t$ and the application $\hat{l}_U$ to $(\mathcal{U}^a_{\exp tU}f)(\theta)$ are equal, for all $f\in L^2(S^1)$.  Moreover, $(\mathcal{U}^a_{\exp tU}f)(\theta)=f(\theta)$ when $t=0$.  Hence, $(\mathcal{U}^a_{\exp tU}f)(\theta)$ is the unique solution to the Cauchy problem
	\begin{equation*}
		\left\{\begin{array}{r c l}
			\displaystyle\frac{d}{dt}S(t,\theta)&=&\hat{l}_US(t,\theta) \\
			S(0,\theta)&=&\mathrm{Id}.
		\end{array}\right.
	\end{equation*}	  	
This means that $\exp(\hat{l}_U)f(\theta)=(\mathcal{U}^a_{\exp U}f)(\theta)$.

\subsection{Representations associated to the trivial orbits}

When $F=(\mu,0)$, the coadjoint orbit of $M(2)$ is the $0$-dimensional
	\begin{equation*}
		\Omega_F=\{(\mu,0)\}
	\end{equation*}
which is a point.  The set of $C^\infty$-functions on this orbit can be described as	
	\begin{equation*}
		C^\infty(\Omega_F)=\{f:\Omega_F\rightarrow\mathbb{C}: f(\mu,0)=z\}\simeq \mathbb{C}.
	\end{equation*}
The Hamiltonian function $\tilde{U}:\Omega_F\rightarrow\mathbb{R}$ is the constant function $\tilde{U}(F)=c_1\mu$.  Obviously, the vector field $\xi_U$ associated to this function is the zero vector field.   The Kirillov form is computed as $\left<F,[U,T]\right>=0$ for any $U,T\in\mathfrak{m}(2)$.

The Moyal $\star$-product on the space $C^\infty(\Omega_F)$ is 
	\begin{equation*}
		f\star g=fg
	\end{equation*}
for any functions $f,g\in C^\infty(\Omega_F)$.  Hence, this $\star$-product is trivially covariant satisfying
	\begin{equation*}
		i\tilde{U}\star i\tilde{T}-i\tilde{T}\star i\tilde{U}=i\widetilde{[U,T]}=0
	\end{equation*}
for any $U,T\in\mathfrak{m}(2)$.  So, there exists a $1$-dimensional representation $l$ of $\mathfrak{m}(2)$ on $C^\infty(\Omega_F))[[\lambda]]$
defined by $(l_U)(f)=i\tilde{U}\star f=(ic_1\mu)f$.  The operator $l_U=0$ when $U\in\mathrm{span}\{E_1,E_2\}$.  

The $1$-parameter subgroup $U=tX,t\in\mathbb{R}$ is identified with $\mathfrak{so}(2)\simeq\mathbb{R}$.  So, the unitary operator $\chi_\mu(\exp tX)=e^{it\mu}$ is the unique solution to the Cauchy problem
	\begin{equation*}
		\left\{\begin{array}{r c l}
			\displaystyle\frac{d}{dt}S(t,x)&=&l_XS(t,x) \\
			S(0,x)&=&\mathrm{Id}
		\end{array}\right.
	\end{equation*}
satisfying $\exp(tl_X)=\chi_\mu(\exp tX)$.

Since the set
	\begin{equation*}
		\{\chi_n\circ p:n\in\mathbb{Z}\}
	\end{equation*}
are the $1$-dimensional UIRs of $M(2)$, the set of orbits  
	\begin{equation*}
		\{\Omega_F=\{(\mu,0)\}:\mu\in\mathbb{Z}\}
	\end{equation*}
corresponds with these $1$-dimensional UIRs and the rest of the non-integer orbits correspond with 
	\begin{equation*}
		\{\chi_\mu\circ p:\mu\in\mathbb{R}/\mathbb{Z}\}.
	\end{equation*}

\section{Conclusion}\label{Conclusion}
This article has aimed to introduce deformation quantization as a powerful tool in constructing and classifying Lie group representations, an alternative to the traditional method of induced representations.  The covariance property of the Moyal $\star-$product and its convergence in the Schwartz space are the key properties that made these constructions and classifications possible.  

The main result of this work is the unitary representation $\mathcal{U}^a$ of $M(2)$.  We have tested Arnal and Cortet's program in \cite{Arnal1,Arnal2,Arnal3}, despite the original design for nilpotent and exponential Lie groups.  The results in section \ref{UIRM(2)} are summarized as follows.
\begin{enumerate}
	\item[1.] The representation $\hat{l}$ of $\mathfrak{m}(2)$ defined by (\ref{DifferentialOperatorsM(2)Corollary}) is the differential representation of the infinite-dimensional UIR of $M(2)$.  Moreover, there is a one-to-one correspondence between the nontrivial orbits and the principal series of UIRs of $M(2)$ and this correspondence is defined by the radius of the cylinder.
	\item[2.] The representation $l$ associated to the orbit $\{(\mu,0)\}$ is the differential representation of the $1$-dimensional unitary irreducible representation of $M(2)$ if $\mu\in\mathbb{Z}$. 
\end{enumerate}

Though the computations in \cite{Diep1,Diep2,Nguyen,Nable} has provided a better understanding of the implementation of the program, this paper implemented it on a cylinder, different from the computations presented in \cite{Arnal8}, and on trivial orbits which was neglected in \cite{Diep1}.  While the program has been effectively implemented on a flat orbit generated by the coadjoint action of a solvable Lie group, it is interesting to extend the said program to an orbit with nonzero curvature generated by a nonsolvable Lie group, for example, the spheres and tangent spheres- these nontrivial orbits are generated by the coadjoint action of $M(3)$.

\acknowledgments

This work was supported by the Commission on Higher Education Faculty Development Program (CHED-FDP) II of the Philippines.


\begin{thebibliography}{99}

\bibitem{Neumann}
J. von Neumann, \emph{Mathematical Foundations of Quantum Mechanics},  translated from the German edition by R. Beyer, Princeton University Press, New Jersey U.S.A. (1955).

\bibitem{Mackey}
G. Mackey, \emph{Harmonic analysis as the exploitation of symmetry- a historical survey},\emph{ Bull. Amer. Math. Soc.} {\bf 3} (1980) 543.

\bibitem{Kirillov2}
A.A. Kirillov, \emph{Lectures on the Orbit Method}, American Mathematical Society, Rhode Island U.S.A. (2004).

\bibitem{Kostant}
B. Kostant, \emph{Quantization and Unitary Representation}, in \text{Lectures in Modern Analysis and Applications III}, R.M. Dudley, J. Feldman, B. Kostant, R.P. Langlands and E.M. Stein, Springer-Verlag, Berlin-Heidelberg (1970), pp. 87-208. 

\bibitem{Souriau}
J. -M. Souriau, \emph{Structure des Syst\'{e}mes Dynamiques},  Dunod, Paris France (1970).

\bibitem{Zachos}
C. Zachos, D. Fairlie and T. Curtright, \emph{Quantum Mechanics in Phase Space}, World Scientific Publishing Co., Singapore (2005).

\bibitem{Wigner}
E. Wigner, \emph{On the quantum correction for thermodynamic equilibrium}, \emph{Phys. Rev.} {\bf 40} (1932) 749.

\bibitem{Weyl}
H. Weyl, \emph{The Theory of Groups and Quantum Mechanics}, Dover Publication, Inc., New York U.S.A. (1931).

\bibitem{Hug}
M. Hug, C. Menke and W.P. Schleich, \emph{Modified spectral method in phase space: calculation of the Wigner function. I. Fundamentals}, \emph{Phys. Rev.} {\bf A 57} (1998) 3188.

\bibitem{Groenewold}
H. Groenewold, \emph{On the principles of elementary quantum mechanics}, \emph{Physica} {\bf 45} (1949) 99.

\bibitem{Moyal}
J.E. Moyal, \emph{Quantum mechanics as a statistical theory}, \emph{Proc. Camb. Phil. Soc.} {\bf 45} (1949) 99.

\bibitem{Bayen}
F. Bayen, M. Flato, C. Fronsdal, A. Lichnerowicz and D. Sternheimer, \emph{Deformation theory and quantization. I. Deformations of symplectic structures}, \emph{Ann. Phys.} {\bf 111} (1978) 61. 

\bibitem{Bayen1}
F. Bayen, M. Flato, C. Fronsdal, A. Lichnerowicz and D. Sternheimer, \emph{Deformation theory and quantization. II. Physical applications}, \emph{Ann. Phys.} {\bf 111} (1978) 111. 

\bibitem{Kontsevich}
M. Kontsevich, \emph{Deformation quantization of Poisson manifolds}, \emph{Lett. Math. Phys.}  {\bf 66} (2003) 157.

\bibitem{Fedosov}
B. Fedosov, \emph{Formal Quantization}, in \emph{Some Topics of Modern Mathematics and Their Application to Problems of Mathematical Physics}, Moscow (1985), pp. 129-136.

\bibitem{Fedosov1}
B. Fedosov, \emph{A simple geometrical construction of deformation quantization}, \emph{J. Differ. Geom.} {\bf 40} (1994) 213. 

\bibitem{Arnal1} 
D. Arnal, \emph{$\star$ products and representations of nilpotent groups}, \emph{Pac. J. Math.} {\bf 114} (1984) 285.

\bibitem{Arnal2} 
D. Arnal and J.C. Cortet, \emph{$\star$-products in the method of orbits for nilpotent groups}, \emph{J. Geom. Phys.} {\bf 2} (1985) 83.

\bibitem{Arnal3}
D. Arnal and J.C. Cortet, \emph{Repr\'esentations $\star$ des groupes exponentiels}, \emph{J. Funct. Anal.} {\bf 82} (1990) 103.


\bibitem{Arnal4} 
D. Arnal, M. Cahen and S. Gutt, \emph{Representations of compact Lie groups and quantization by deformation}, \emph{Bull. Acad. Royale Belg.} {\bf 74} (1988) 123.

\bibitem{Arnal7} 
D. Arnal, J.C. Cortet and J. Ludwig, \emph{Moyal product and representations of solvable Lie groups}, \emph{J. Funct. Anal.} {\bf 133} (1995) 402.

\bibitem{Arnal5}
D. Arnal, J.C. Cortet, P. Molin and G. Pinczon, \emph{Covariance and geometrical invariance in quantization}, \emph{Lett. Math. Phys.} {\bf 24} (1983) 276.

\bibitem{Fronsdal}
C. Fronsdal, \emph{Some ideas about quantization}, \emph{Rep. Math. Phys.} {\bf 15} (1978) 111.

\bibitem{Moreno}
C. Moreno, \emph{Invariant star products and representations of compact semisimple Lie groups}, \emph{Lett. Math. Phys.} {\bf 12} (1986) 217.

\bibitem{Hirshfeld}
A. Hirshfeld and P. Henselder, \emph{Deformation quantization in the teaching of quantum mechanics}, \emph{Am. J. Phys.} {\bf 70} (2002) 537.

\bibitem{Case}
W. Case, \emph{Wigner's functions and Weyl transforms for pedestrians}, \emph{Am. J. Phys.} {\bf 76} (2008) 937.

\bibitem{Diep1}
Do Ngoc Diep and Nguyen Viet Hai, \emph{Quantum half-planes via deformation quantization}, \emph{Beitr. Algebra Geom.} {\bf 42} (2001) 407.

\bibitem{Diep2}
Do Ngoc Diep and Nguyen Viet Hai, \emph{Quantum co-adjoint orbits of the group of affine transformation of the complex line}, \emph{Beitr. Algebra Geom.} {\bf 42} (2001) 419.

\bibitem{Nable}
J. Nable, \emph{Deformation quantization and representations of the real rotation group}, \emph{Science Diliman} {\bf 13} (2001) 41.

\bibitem{Nguyen}
Nguyen Viet Hai, \emph{Quantum co-adjoint orbits of $MD_4$-groups}, \emph{Vietnam J. Math.} {\bf 29} (2001) 131.

\bibitem{Berndt}
R. Berndt, \emph{An Introduction to Symplectic Geometry}, American Mathematical Society, Rhode Island U.S.A. (2000).

\bibitem{Abraham}
R. Abraham and J. Marsden, \emph{Foundations of Mechanics 2nd ed.}, Addison-Wesley Publishing, Inc., Canada (1978).

\bibitem{Sugiura}
M. Sugiura, \emph{Unitary Representations and Harmonic Analysis- An Introduction}, North-Holland, Amsterdam-Oxford-New York-Tokyo (1990).

\bibitem{Berndt1}
R. Berndt, \emph{Representations of Linear Groups}, Friedr. Vieweg \& Sohn Verlag, Berlin, Germany (2007).

\bibitem{Mackey1}
G. Mackey, \emph{Theory of Unitary Group Representations}, The University of Chicago Press, London U.K. (1976).

\bibitem{Vilenkin2}
N. Ja. Vilenkin and A.U. Klimyk, \emph{Representations of Lie Groups and Special Functions, Volume 1: Simplest Lie groups, special functions and integral transforms},  Kluwer Academic Publishers, Dordrecht The Netherlands (1991).

\bibitem{Vilenkin3}
N. Ja. Vilenkin and A.U. Klimyk, \emph{Representations of Lie groups and special functions, Volume 2: Class I representations, special functions, and integral transforms},  Kluwer Academic Publishers, Dordrecht The Netherlands (1991).

\bibitem{Vilenkin4}
N. Ja. Vilenkin and A.U. Klimyk, \emph{Representations of Lie groups and special functions, Volume 3: Classical and quantum groups and special functions},  Kluwer Academic Publishers, Dordrecht The Netherlands (1991).

\bibitem{Vilenkin}
N. Ja. Vilenkin, \emph{Special Functions and the Theory of Group Representations}, translated from Russian by V.N. Singh, American Mathematical Society, Rhode Island U.S.A. (1968).

\bibitem{Mackey3}
G. Mackey, \emph{On induced representations of groups}, \emph{Am. J. Math.} {\bf 73} (1951) 576.

\bibitem{Mackey2}
G. Mackey, \emph{Induced representations of locally compact groups I}, \emph{Ann. Math.} {\bf 55} (1952) 101.


\bibitem{Kirillov}
A.A. Kirillov \emph{Geometric quantization}, in \emph{Dynamical Systems IV}, V.I. Arnol'd and S.P Novikov, Springer-Verlag, New York, Berlin, Heidelberg (1985) pp. 137-172.

\bibitem{Kirillov1}
A.A. Kirillov, \emph{Unitary representations of nilpotent Lie groups} (Russian),  \emph{Uspekhi Mat. Nauk}  {\bf 17} (1962) 57.

\bibitem{Kirillov3}
A.A. Kirillov, \emph{Introduction to the theory of representations and noncommutative harmonic analysis}, in \emph{Representation Theory and Noncommutative Harmonic Analysis I}, A.A. Kirillov, Springer-Verlag  New York, Berlin, Heidelberg (1991) pp. 1-156.

\bibitem{Dirac}
P. Dirac, \emph{The Principles of Quantum Mechanics, 4th ed.} Clarendon Press, Oxford (1957).

\bibitem{Gutt}
S. Gutt, \emph{Deformation quantization}, in \textit{Workshop on Representation Theory of Lie Groups}, International Center for Theoretical Physics, SMR.686/14.

\bibitem{Bordemann}
M. Bordemann, \emph{Deformation quantization: a survey}, \emph{J. Phys.} {\bf 103} (2008) 1. 

\bibitem{Hansen}
F. Hansen, \emph{Quantum mechanics in phase space},  \emph{Rep. Math. Phys.} {\bf 19} (1984) 361.

\bibitem{Arnal8}
D. Arnal and J.C. Cortet, \emph{Star representations of $E(2)$}, \emph{Lett. Math. Phys.} {\bf 20} (1990) 141.

\end{thebibliography}
\end{document}